\documentclass{ws-procs975x65}

\usepackage[T1]{fontenc}
\usepackage[latin9]{inputenc}
\usepackage{mathrsfs}

\usepackage{esint}
\usepackage{bm}
\usepackage{bbm}

\usepackage{hyperref}

\renewcommand\[{\begin{equation}}
\renewcommand\]{\end{equation}}

\newcommand{\ra}{\rightarrow}

\def\d{\mbox d}

\def\com{}
\def\cob{}
\newcommand{\oarX}[1]{\href{http://arxiv.org/abs/#1}{{\ttfamily \com #1}}}
\newcommand{\arX}[1]{\href{http://arxiv.org/abs/#1}{{\ttfamily \com arXiv:#1}}}
\newcommand{\doin}[5]{\href{http://dx.doi.org/#1}{\cob \textit{#2} \textbf{#3} (#5) #4}}
\newcommand{\ndoin}[5]{\href{#1}{\cob \textit{#2} \textbf{#3} (#5) #4}}

\begin{document}

\title{\uppercase{Fields and Laplacians on Quantum Geometries}}

\author{\uppercase{Johannes Th\"urigen}}

\address{Max Planck Institute for Gravitational Physics (Albert Einstein Institute)\\
Am M\"uhlenberg 1, D-14476 Potsdam, Germany}

\begin{abstract}
In fundamentally discrete approaches to quantum gravity such as loop quantum gravity, spin-foam models, group field theories or Regge calculus observables are functions on discrete geometries. We present a bra-ket formalism of function spaces and discrete calculus on abstract simplicial complexes equipped with geometry and apply it to the mentioned theories of quantum gravity. In particular we focus on the quantum geometric Laplacian and discuss as an example the expectation value of the heat kernel trace from which the spectral dimension follows.
\end{abstract}


\bodymatter

\section{Introduction and overview}

Despite a wealth of recent results an outstanding challenge in current approaches to quantum gravity such as loop quantum gravity (LQG) \cite{Rovelli:2004wb}
and spin-foam models \cite{Perez:2012uz}, group field theory\cite{GFT2}, quantum Regge calculus\cite{Hamber:2009zz} or (causal) dynamical triangulations (CDT)  \cite{Ambjorn:2010kv} is to recover geometric information from their discrete building blocks of geometry and spacetime, 
either understood as fundamentally discrete or merely as a regularization to define the theory.

One example of geometric observable that has been widely used as a probe of the geometry of states, phases or histories
\cite{AJL4,LaR5,Mod08,BeH,frc4}
is the spectral dimension $d_s$,
capturing its diffusion properties via the trace of the heat kernel $K_{\sigma \sigma'}(\tau)$, formally 
\[
\left\langle \widehat{P}(\tau)\right\rangle =\left\langle \mbox{Tr}\widehat{K}_{\sigma \sigma'}(\tau)\right\rangle=\left\langle \mbox{Tr}e^{\tau\widehat{\Delta}}\right\rangle \sim\tau^{\frac{d_{s}}{2}}.
\]
It depends on the underlying geometry through the Laplacian operator $\Delta$, thereby implicitly relying on some notion of a (test) matter field it is acting on.

In this  contribution we present a formalism for rigorously defining Laplacians on discrete and quantum geometries (any details can be found in  Ref.~\refcite{COT1}). This is needed for the analysis of the spectral dimension in LQG, spin-foams and group field theory \cite{COT2}.
In a first step we introduce a framework for the definition of $p$-form fields and differential operators based on the recently developed {\it discrete calculus} of Ref.~\refcite{Desbrun:2005ug}, but generalized to finite abstract simplicial complexes endowed with geometric data, and we discuss the properties of the corresponding Laplacian. 
This sets the stage to define, in a second step, (functions of) the Laplacian as quantum observables in terms of various geometric variables used in the approaches to quantum gravity.


\section{Laplacian on simplicial pseudo-manifolds}

To be applicable to all models of quantum geometry we choose the most general setting of a finite abstract simplicial $d$-complex $K$. 
It shall fulfill pseudo-manifold properties 
\cite{Gurau:2010iu} 
to allow for the definition of a dual complex $\star K$.
Finally, for a $p \le d$, an assignment of 
$p$-volumes $V_{\sigma_{p}}$,
dual $(d-p)$-volumes $V_{\star \sigma_{p}}$ and
support $d$-volumes $V_{\sigma_{p}}^{(d)}$
to every primal $p$-simplex $\sigma_p\in K$ and its dual $\hat\sigma_{d-p} := \star \sigma_p\in \star K$ is needed.

Now a $p$-form field $\phi$ is a $p$-cochain on $K$ and has a Hodge dual $(d-p)$-form defined as the corresponding form on the dual complex $\star K$ \cite{Desbrun:2005ug}.
By identifying the primal chain basis elements with the dual cochain basis elements we can merge the chain-cochain duality with the duality between primal and dual complex. 
This gives rise to a unique orthonormal and complete position basis, in bra-ket notation and in a convention explicitly keeping track of the position space measure\cite{COT1}
\[
\left\langle \sigma_{p}|\sigma'_{p}\right\rangle =\frac{1}{V_{\sigma_{p}}^{(d)}}\delta_{\sigma\sigma'}\\
\ \ ,\ \ \ \\
\underset{\sigma_{p}}{\sum}V_{\sigma_{p}}^{(d)}|\sigma_{p}\rangle\langle\sigma_{p}|=\mathbbm{1}.
\]
Primal and Hodge dual field can then be expanded in position coefficients as
\[
\langle\phi|=\underset{\sigma_{p}\in K}{\sum}V_{\sigma_{p}}^{(d)}\phi_{\sigma_{p}}\langle\sigma_{p}|\overset{*}{\longleftrightarrow}|\phi\rangle=\underset{\sigma_{p}\in K}{\sum}V_{\sigma_{p}}^{(d)}\phi_{\sigma_{p}}^{*}|\sigma_{p}\rangle.
\]
Taking Stokes theorem as definition, the differential $\d$ is defined in terms of the boundary operator on chains
with an analogous definition for the adjoint differential $\delta$ in terms of the differential on the dual.\cite{Desbrun:2005ug,COT1}
Consequently this defines the usual Hodge-Laplacian $\Delta=\d\delta+\delta\d$. On a dual scalar field
\[
\left(-\Delta\phi\right)_{\hat{\sigma}_{0}}=\frac{1}{V_{\sigma_{d}}}\ \underset{\sigma'_{d}\sim\sigma_{d}}{\sum}\frac{V_{\sigma_{d}\cap\sigma_{d}'}}{V_{\star\left(\sigma_{d}\cap\sigma_{d}'\right)}}\left(\phi_{\hat{\sigma}_{0}}-\phi_{\hat{\sigma}'_{0}}\right)\label{lap}.
\]
By definition this Laplacian fulfills a null condition, self-adjointness and locality 
and we have shown a naive convergence to the continuum Laplacian under refinement in the case of triangulations of smooth manifolds. 
In that case it makes a difference whether one chooses volumes in terms of a circumcentric or barycentric dual.
Due to the possibility of circumcenters lying outside their simplex, in general only the latter guarantees positivity of the Laplacian coefficients
(and consequently, OS positivity).

In the case of only finite volume factors  the eigenfunctions of the Laplacian $e_{\sigma}^{\lambda}=\left\langle \lambda|\sigma\right\rangle $
form an orthonormal and complete momentum basis giving rise to a momentum transform in which functions of $\Delta$ can be expanded, e.g. the heat trace
\[
K_{\sigma\sigma'}(\tau)=\left\langle \sigma'|e^{\tau\Delta}|\sigma\right\rangle =\underset{\lambda}{\sum}\frac{1}{V_{\lambda}}e^{-\tau\lambda}e_{\sigma'}^{\lambda*}e_{\sigma}^{\lambda}
\ \ra \ 
P(\tau)=\mbox{Tr}K_{\sigma\sigma'}(\tau)=\frac{1}{V}\underset{\lambda}{\sum}e^{-\tau\lambda}\label{heat}
\]

Note that the formalism can be made sufficiently general to be extended to polyhedral complexes and complexes with boundary as well \cite{COT1}.


\section{Laplacian in models of quantum geometry}

The discrete Laplacian depends both on the combinatorial structure of the complex and on the discrete geometry via the various volume factors. As a first step towards quantum geometries we have constructed the latter in terms of the geometric variables used in various models of quantum geometry, i.e. in edge lengths, face normals, area-angle variables and bivector/flux variables.\cite{COT1}

For their quantization, either canonical or via a path integral, a main challenge is to deal with possible singularities of the matrix entries of the Laplacian, coming from inverse volume factors.
In a canonical setting, these singularities may prevent the definition of the Laplacian operator as a bounded operator; in the covariant setting, they may produce divergences in explicit evaluations.
Whether or not such difficulties arise depends on the details of the quantum theory considered, 
such as the precise structure of the Hilbert space of states or the path integral measure, and on the exact classical expression to be quantized.
Furthermore, for many purposes, it is not the Laplacian as such but functions of it which are of interest.
These need not have the same quantization issues as the Laplacian itself.
One would expect, for example, that the heat trace (eq. \ref{heat}) vanishes exactly when the Laplacian is singular.
Thus one may even envisage cases in which observables as functions of the Laplacian inserted within quantum geometric evaluations might help to suppress pathological configurations corresponding to degenerate or divergent geometries.


\section{Conclusion and Outlook}

Based on discrete calculus we have presented a formalism for differential operators and arbitrary fields on discrete pseudo-manifolds in which observables like the heat trace have a well-defined meaning, and applied it to various models of quantum gravity. This should open up novel ways to investigate the physical and geometric properties of these models. 

Also, the Laplacian enters the definition of an invertible momentum transform to a representation of fields on its eigenspaces. This generalization of the Fourier transform can be effectively used to handle functions of the Laplacian such as the spectral dimension of spacetime\cite{COT2}. 
Another application would be as a necessary ingredient for defining matter coupling in discrete models of quantum gravity.



\begin{thebibliography}{9}

\bibitem{Desbrun:2005ug}
M.~Desbrun, A.N.\ Hirani, M.~Leok, and J.E.\ Marsden,
\oarX{math/0508341}.

\bibitem{Rovelli:2004wb}
C.~Rovelli, {\em {Quantum Gravity}}, Cambridge University Press, Cambridge U.K. (2004).

\bibitem{Perez:2012uz}
A.~P{\'e}rez,
 \ndoin{http://relativity.livingreviews.org/Articles/lrr-2013-3/}{Liv.~Rev.\ Rel.}{16}{3}{2013}.

\bibitem{GFT2} D.~Oriti,
2012 \emph{Foundations of Space and Time: Reflections on Quantum Gravity} ed G Ellis \emph{et al} (Cambridge:
Cambridge University Press) [\arX{1110.5606}].

\bibitem{Hamber:2009zz}
H.W.\ Hamber, {\em {Quantum Gravitation: The Feynman Path Integral Approach}}, Springer, Amsterdam The Netherlands (2008).

\bibitem{Ambjorn:2010kv} J.~Ambj{\o}rn, J.~Jurkiewicz, and R.~Loll,
\doin{10.1007/978-3-642-11897-5_2 }{Lect.\ Notes\ Phys.}{807}{59}{2010} 

\bibitem{AJL4}  J.~Ambj{\o}rn, J.~Jurkiewicz, and R.~Loll,
\doin{10.1103/PhysRevLett.95.171301}{Phys.\ Rev.\ Lett.}{95}{171301}{2005}

\bibitem{LaR5} 
O.~Lauscher and M.~Reuter,
\doin{10.1088/1126-6708/2005/10/050}{JHEP}{10}{050}{2005} 
[\oarX{hep-th/0508202}].

\bibitem{Mod08} 
L.~Modesto,
\doin{10.1088/0264-9381/26/24/242002}{Class.\ Quantum\ Grav.}{26}{242002}{2009}  
[\arX{0812.2214}].

\bibitem{BeH}   D.~Benedetti and J.~Henson, 
\doin{10.1103/PhysRevD.80.124036}{Phys.\ Rev.\ D}{80}{124036}{2009} [\arX{0911.0401}].

\bibitem{frc4}  Calcagni G
\doin{10.1103/PhysRevE.87.012123}{Phys.\ Rev.}{E}{ 87}{012123}{2013} [\arX{1205.5046}].

\bibitem{COT1}  G.\ Calcagni, D.\ Oriti, and J.\ Th\"urigen,
\arX{1208.0354}

\bibitem{COT2}  G.\ Calcagni, D.\ Oriti, and J.\ Th\"urigen, in progress.

\bibitem{Gurau:2010iu}
R.~Gurau,
 \doin{10.1088/0264-9381/27/23/235023}{Class.\ Quantum\ Grav.} {27} {235023}{2010}
  [\arX{1006.0714}].
  

\end{thebibliography}
\end{document}